\documentclass[conference]{IEEEtran}

\usepackage[T1]{fontenc}
\usepackage[utf8]{inputenc}
\usepackage{amssymb}
\usepackage[pdftex]{graphicx}
\usepackage{subcaption} 
\usepackage{amsmath}
\usepackage{amsthm}
\usepackage{mathtools}
\usepackage[noadjust]{cite}
\usepackage{fullpage}
\usepackage[hidelinks]{hyperref}
\usepackage[acronym]{glossaries}
\usepackage{siunitx}
\usepackage{tikz}
\usepackage{flushend}
\usepackage{multirow}

\newacronym{NPRACH}{NPRACH}{narrowband physical random-access channel}
\newacronym{ToA}{ToA}{time of arrival}
\newacronym{CFO}{CFO}{carrier frequency offset}
\newacronym{NBIoT}{NB-IoT}{narrowband internet of things}
\newacronym{5GNR}{5G NR}{5G New Radio}
\newacronym{3GPP}{3GPP}{3rd Generation Partnership Project}
\newacronym{UMi}{UMi}{urban microcell}
\newacronym{RMSE}{RMSE}{root-mean-square error}
\newacronym{NN}{NN}{neural network}
\newacronym{BS}{BS}{base station}
\newacronym{UE}{UE}{user equipment}
\newacronym{SG}{SG}{symbol group}
\newacronym{CP}{CP}{cyclic prefix}
\newacronym{OFDM}{OFDM}{orthogonal frequency division multiplexing}
\newacronym{FFT}{FFT}{fast Fourier transform}
\newacronym{AWGN}{AWGN}{additive white Gaussian noise}
\newacronym{DFT}{DFT}{discrete Fourier transform}
\newacronym{FNR}{FNR}{false negative rate}
\newacronym{FPR}{FPR}{false positive rate}
\newacronym{RG}{RG}{resource grid}
\newacronym{RE}{RE}{resource element}
\newacronym{SNR}{SNR}{signal-to-noise ratio}
\newacronym{1D}{1D}{one-dimensional}
\newacronym{MLP}{MLP}{multilayer perceptron}
\newacronym{KL}{KL}{Kullback–Leibler}
\newacronym{SGD}{SGD}{stochastic gradient descent}
\newacronym{ppm}{ppm}{parts-per-million}
\newacronym{ICI}{ICI}{inter-carrier interference}

\renewcommand{\vec}[1]{\mathbf{#1}}

\newcommand{\yv}{\vec{y}}

\newcommand{\Ym}{\vec{Y}}
\newcommand{\Zm}{\vec{Z}}

\newcommand{\Fc}{{\cal F}}

\newcommand{\Lc}{{\cal L}}

\newcommand{\CC}{\mathbb{C}}

\newcommand{\RR}{\mathbb{R}}

\newcommand{\LB}{\left(}
\newcommand{\RB}{\right)}
\newcommand{\LP}{\left\{}
\newcommand{\RP}{\right\}}
\newcommand{\LSB}{\left[}
\newcommand{\RSB}{\right]}

\renewcommand{\ln}[1]{\mathop{\mathrm{ln}}\LB #1\RB}

\newcommand{\EE}{{\mathbb{E}}}

\newcommand\abs[1]{\left| #1 \right|}

\begin{document}

\title{Deep Learning-Based Synchronization for Uplink NB-IoT}

\author{\IEEEauthorblockN{Fay\c{c}al A\"{i}t Aoudia, Jakob Hoydis, Sebastian Cammerer,\\
Matthijs Van Keirsbilck, and Alexander Keller}
		\IEEEauthorblockA{NVIDIA\\Contact: faitaoudia@nvidia.com}
}

\maketitle

\begin{abstract}
We propose a \gls{NN}-based algorithm for device detection and \gls{ToA} and \gls{CFO} estimation for the \gls{NPRACH} of \gls{NBIoT}.
The introduced \gls{NN} architecture leverages residual convolutional networks as well as knowledge of the preamble structure of the \gls{5GNR} specifications.
Benchmarking on a \gls{3GPP} \gls{UMi} channel model with random drops of users against a state-of-the-art baseline shows that the proposed method enables up to \SI{8}{\decibel} gains in \gls{FNR} as well as significant gains in \gls{FPR} and \gls{ToA} and \gls{CFO} estimation accuracy.
Moreover, our simulations indicate that the proposed algorithm enables gains over a wide range of channel conditions, \glspl{CFO}, and transmission probabilities.
The introduced synchronization method operates at the \gls{BS} and, therefore, introduces no additional complexity on the user devices.
It could lead to an extension of battery lifetime by reducing the preamble length or the transmit power.
Our code is available at: \url{https://github.com/NVlabs/nprach_synch/}.
\end{abstract}

\glsresetall

\section{Introduction}

\Gls{NBIoT} is a radio technology standard developed by the \gls{3GPP} to enable a wide range of IoT services and \glspl{UE}~\cite{9194757}.
It mostly reuses \gls{5GNR} specifications, and aims to support massive numbers of connected \glspl{UE} and to provide very good outdoor-to-indoor coverage, very low power consumption to enable long battery lifetime, and low cost connectivity.
In the uplink, many \glspl{UE} can simultaneously contend to access the channel in a random access manner.
The \gls{NPRACH} relates to the first message sent by a \gls{UE} to a \gls{BS} to request access to the channel, and is used by the \gls{BS} to identify the \gls{UE} and estimate its \gls{ToA} and \gls{CFO}.

The \gls{NPRACH} waveform~\cite{ts36211} is specified as a single-tone frequency hopping preamble, and different configurations are available to adapt to various channel conditions and cell sizes.
Frequency hopping is performed according to a pseudo-random pattern to mitigate inter-cell interference as neighboring cells use different hopping patterns.
Within a cell, up to 48 orthogonal hopping patterns are available for the \glspl{UE} to choose from, and the \glspl{UE} that simultaneously request access to the channel must use different patterns to avoid collisions.
The problem of \gls{NPRACH} detection consists in jointly detecting the \glspl{UE} that simultaneously attempt to access the channel and estimating their respective \glspl{ToA} and \glspl{CFO}.

To make the problem tractable, many existing algorithms~\cite{7569029,8922625,9263250} assume that (i) the channel frequency response is flat, which is reasonable considering the narrow bandwidth of \gls{NBIoT} (\SI{180}{\kilo\Hz}), (ii) interference between \glspl{UE} can be neglected, which only holds true assuming low \gls{CFO} and low mobility, and (iii) the channel is time-invariant over the duration of the preamble, which is only valid under low mobility.
Moreover, most algorithms require the configuration of a detection power threshold that depends on the noise-plus-interference level, and which controls the trade-off between the occurrence of false positives and false negatives.

We propose a deep learning-based synchronization algorithm that requires none of these assumptions to hold, as well as no configuration of a detection threshold.
The proposed algorithm is standard-compliant and operates at the \gls{BS}, leading to no additional complexity for the \glspl{UE}. 
To the best of our knowledge, the only related work is~\cite{9013510}, which uses a convolutional neural network to predict the active \glspl{UE} and corresponding \glspl{ToA} and \glspl{CFO}.
In comparison, the method we propose uses a different \gls{NN} architecture which exploits knowledge of the preamble structure to achieve increased performance.
Moreover, it uses a different loss function for training which is key to enable accurate \gls{ToA} and \gls{CFO} estimates.

\begin{figure*}
	\center
	\includegraphics[width=\linewidth]{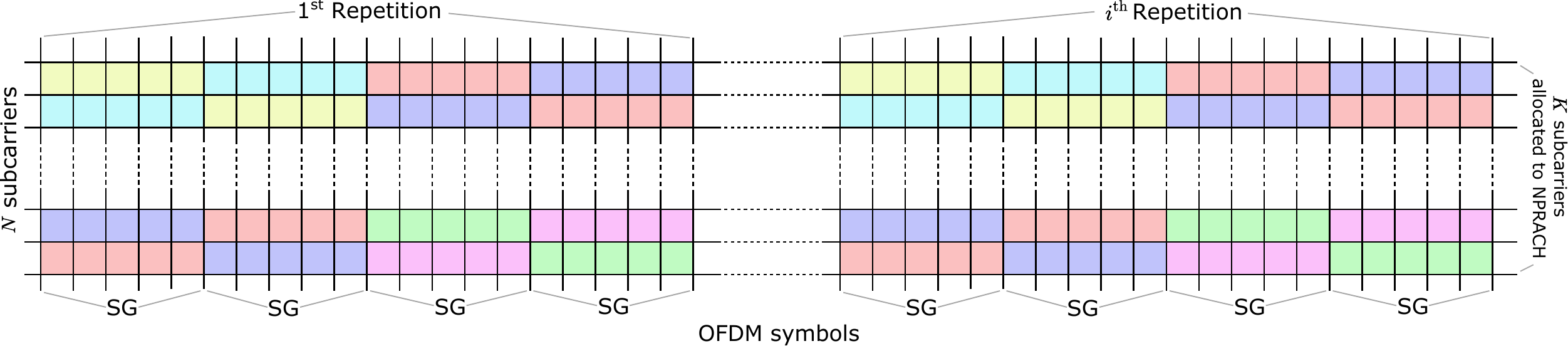}
	\caption{NPRACH structure. Each color correspond to a preamble transmitted by a user.\label{fig:nprach}}
\end{figure*}

Benchmarking using the Sionna link-level simulator~\cite{hoydis2022sionna} on a realistic \gls{3GPP} \gls{UMi} channel model~\cite{ts38901} and against a state-of-the-art baseline~\cite{9263250} shows that the proposed algorithm enables gains of up to \SI{8}{\decibel} in \gls{FNR}, significant reduction in \gls{FPR}, and more accurate \gls{ToA} and \gls{CFO} estimation.
Moreover, these results hold over a wide range of \glspl{CFO} and transmission probabilities.
As these gains were obtained for a short preamble length, such an algorithm may remove the need for longer preambles under many channel conditions or reduce the required transmit power, leading to battery lifetime extension which is critical for \gls{NBIoT} applications.

\section{System Model}

The \gls{NPRACH} waveform~\cite{ts36211} consists of a sequence of \glspl{SG}, as shown in Fig.~\ref{fig:nprach}.
Each \gls{SG} is made of five identical single-tone \gls{OFDM} symbols that share a single \gls{CP} to reduce overhead, and occupies one tone of \SI{3.75}{\kilo\Hz} bandwidth.
Frequency hopping is performed between the \glspl{SG}, and four consecutive \glspl{SG} are treated as the basic unit of the preamble, referred to as a \emph{repetition}.
A preamble can consist of up to 128 repetitions for coverage extension.

Let us denote by $K$ the maximum number of devices that can simultaneously access the channel.
It is assumed that \glspl{UE} that simultaneously request access do not collide, i.e., that they use different hopping patterns.
Some methods in the literature address the detection of colliding \glspl{UE}, e.g.,~\cite{9449922}.
However, there is currently no procedure to acknowledge access requests from colliding \glspl{UE}.
Under this assumption, the highest value allowed for $K$ is 48, as this is the highest number of hopping patterns available~\cite{ts36211}.
The samples transmitted by the $k^{\text{th}}$ user for the $i^{\text{th}}$ symbol of the $m^{\text{th}}$ \gls{SG} is
\begin{equation}
    s_{k,m,i}[n] = \beta_k e^{j 2 \pi \phi_k[m] \frac{n}{N}}
\end{equation}
where $\beta_k$ is the transmission power used by the $k^{\text{th}}$ user, $N$ is the number of subcarriers, and $\phi_k[m]$ is the subcarrier index used by the $k^{\text{th}}$ user for the $m^{\text{th}}$ \gls{SG} and is determined by the hopping pattern. Note that $N$ is typically greater than $K$ as the \gls{NPRACH} is only part of the radio spectrum processed by the \gls{BS}.

We consider a multi-user time-invariant multipath channel, where the channel response of the $k^{\text{th}}$ user is
\begin{equation}
	h_k(\tau) = \sum_{p=0}^{P_k-1} a_{k,p} \delta \LB \tau - \tau_{k,p} \RB
\end{equation}
where $P_k$ is the number of paths for the $k^{\text{th}}$ user, and $a_{k,p}$ and $\tau_{k,p}$ are the baseband coefficient and delay of the $p^{\text{th}}$ path of the $k^{\text{th}}$ user, respectively.
The \gls{ToA} of the $k^{\text{th}}$ user is defined as
\begin{equation}
	D_k \coloneqq \min_{p} \tau_{k,p}.
\end{equation}
The received signal at the \gls{BS} is
\begin{multline}
	y_{m,i}[n] = \sum_{k=0}^{K-1} \sum_{\ell = -\infty}^{\infty} A_k h_{k,\ell} s_{k,m,i} \LSB n-\ell \RSB e^{j 2 \pi f_{\text{off},k} \LB n - \ell \RB}\\
	+ w_{m,i}[n]
\end{multline}
where $A_k$ indicates if user $k$ is active, i.e., it equals $1$ if the user is transmitting and $0$ otherwise, $h_{k,\ell}$ is the channel coefficient of the $\ell^{\text{th}}$ tap and of the $k^{\text{th}}$ user, $f_{\text{off},k}$ is the \gls{CFO} of the $k^{\text{th}}$ user normalized by the sampling frequency, and $w_{m,i}[n]$ is the \gls{AWGN} with variance $\sigma^2$.
To simulate the channel, summation over taps is performed for a finite number of time-lags $ L_{\text{min}} \leq \ell \leq L_{\text{max}}$.
Assuming sinc pulse shaping on the transmitter side and matched filtering on the receiver side, the channel taps are
\begin{align}
	h_{k, \ell} &= \sum_{p=0}^{P_k-1} a_{k,p} \text{sinc}\LB \ell - W\tau_{k,p} \RB\\
				&= \Fc^{-1} \LP \text{rect}\LB f \RB H_k(f) \RP \LB \ell \RB \label{eq:ch_freq}
\end{align}
where $W = N \Delta_f$ is the bandwidth, $\Fc^{-1}\LP X(f) \RP \LB t \RB$ the inverse Fourier transform of $X(f)$ evaluated at $t$, and
\begin{equation}
	H_k(f) = \sum_{p=0}^{P_k-1} a_{k,p} e^{-j2\pi \tau_{k,p} W f}
\end{equation}
the frequency response of the channel.

On the receiver side, the \gls{CP} of each \gls{SG} is removed and \gls{DFT} is performed.
This leads to a \gls{RG} $\Ym$ of size $N \times 5S$ where $S$ is the number of \glspl{SG} forming the preamble.
Considering the $k^{\text{th}}$ user, the received signal for the for the $i^{\text{th}}$ symbol of the $m^{\text{th}}$ \gls{SG} is
\begin{multline}
	\label{eq:rx_freq}
	Y\LSB \phi_k[m], 5m+i \RSB = A_k H_k \LB \frac{\phi_k[m]}{N} \RB \beta_k \cdot\\
	\frac{1}{N} \sum_{n=N_{m,i}}^{N_{m,i}+N-1} e^{j 2 \pi f_{\text{off},k}n} + \sum_{k' \neq k} \Bigg\{ A_{k'} H_{k'} \LB \frac{\phi_k[m]}{N} \RB \beta_{k'} \cdot\\
	\frac{1}{N} \sum_{n=N_{m,i}}^{N_{m,i}+N-1} e^{j 2 \pi \LB \frac{\phi_{k'}[m] - \phi_{k}[m]}{N} + f_{\text{off},k'} \RB n} \Bigg\} \\
		 + W_{k,m,i} \LSB \phi_k[m], 5m+i \RSB
\end{multline}
where $W \LSB \phi_k[m], 5m+i \RSB$ is \gls{AWGN} with variance $\sigma^2$, and $N_{m,i} \coloneqq m N_{\text{SG}} + i N$ where $N_{\text{SG}}$ is the number of samples forming an \gls{SG}.
The hopping patterns are orthogonal in time and frequency, i.e., $\phi_k[m] \neq \phi_{k'}[m]$ when $k \neq k'$.
The first term on the right-hand side of~\eqref{eq:rx_freq} is the signal received from the $k^{\text{th}}$ user, and the second term corresponds to \gls{ICI} from the other users due to \gls{CFO}.
\gls{NPRACH} synchronization consists in jointly detecting the active users and estimating their \gls{ToA} and \gls{CFO} from the received signal $\Ym$~\eqref{eq:rx_freq}.
The next section introduces a \gls{NN}-based detector that aims to achieve this goal by exploiting the \gls{NPRACH} preamble structure.

\section{Deep Learning-based synchronization}

\begin{figure}
	\center
	\includegraphics[width=\columnwidth]{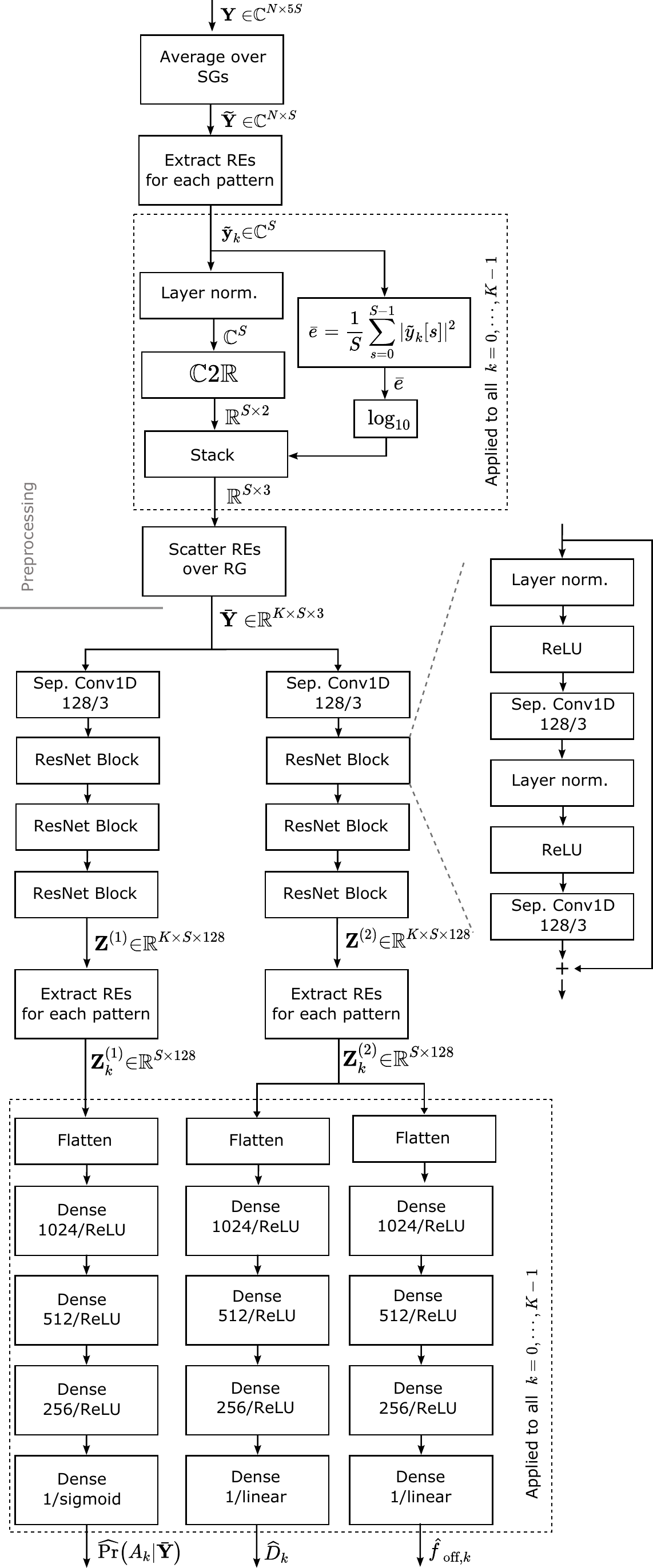}
	\caption{NN-based synchronization algorithm. When labeling the output of a layer is not required, only the shape is indicated. For dense layers, the number of units/activation function is indicated.
	For separable convolution layers, the number of kernels/kernel size is indicated.\label{fig:nn}}
\end{figure}

Fig.~\ref{fig:nn} shows the algorithm that we propose for \gls{NPRACH} synchronization.
It takes as input the received \gls{RG} $\Ym \in \CC^{N \times 5S}$, which is first preprocessed into a real-valued tensor $\bar{\Ym} \in \RR^{K \times S \times 3}$.
$\bar{\Ym}$ then serves as input to two \glspl{NN}, one to compute transmission probabilities, denoted by $\widehat{\Pr} \LB A_k \lvert \bar{\Ym} \RB$, and the other to compute \gls{ToA} and \gls{CFO} estimates, denoted by $\widehat{D}_k$ and $\hat{f}_{\text{off},k}$, respectively.
The rest of this section details the preprocessing, the \gls{NN} architectures, and the loss function used for training.

\subsection{Preprocessing}

Preprocessing consists of first averaging the five \glspl{RE} forming each \gls{SG}.
This is done separately for each subcarrier.
This preprocessing step reduces by a factor of five the input size, resulting in a matrix $\widetilde{\Ym} \in \CC^{N \times S}$.

The second preprocessing step consists of separately normalizing the sequences of \glspl{SG} corresponding to the $K$ possible hopping patterns.
This is key to enable accurate detection despite the users having \glspl{SNR} that differ by orders of magnitude due to path loss.
This is achieved by gathering the $K$ sequences corresponding to every hopping pattern $\tilde{\yv}_k \in \CC^S, 0 \leq k \leq K-1$, according to the \gls{3GPP} specifications~\cite{ts36211}, and normalizing each sequence individually.
The resulting normalized sequences are then converted from complex-valued tensors to real-valued ones by stacking the real and imaginary components along an additional dimension, resulting in a tensor of shape $S \times 2$.
As normalization erases the information about the received power, the average received power of each sequence is computed prior to normalization in log-scale and concatenated to the normalized sequence along the inner dimension.
The received power takes values over a range of several orders of magnitude due to path loss.
Finally, the $K$ resulting tensors are scattered according to the hopping patterns, reverting the previous gathering operation, to form a normalized tensor $\bar{\Ym} \in \RR^{K \times S \times 3}$.
This allows the \glspl{NN} to operate over the time-frequency \gls{RG} and hence to better mitigate the \gls{ICI} due to \gls{CFO}, as opposed to conventional approaches which typically ignore \gls{ICI} for tractability.

\subsection{Neural Network Architecture}

The tensor $\bar{\Ym}$ serves as input to two \glspl{NN} that share a similar architecture.
For each user, the first \gls{NN} computes a probability of the user to request channel access, and the second \gls{NN} computes estimates of the \gls{ToA} and the \gls{CFO} of the user.
The first stage of both \glspl{NN} is motivated by the \gls{ICI} caused by the \gls{CFO}, and consists of \gls{1D} depth-wise separable convolutional layers~\cite{Chollet_2017_CVPR} with 128 kernels of size 3 and that operate along the frequency dimension.
Depth-wise separable convolutional layers are used as they significantly decrease computational complexity compared to conventional convolutional layers without reducing accuracy.
Skip connections are used to avoid gradient vanishing, and zero-padding ensures that the output has the same length as the input.
The resulting tensors are denoted by $\Zm^{(1)}$ and $\Zm^{(2)}$ for the first and second \gls{NN}, respectively, and have shape $K \times S \times 128$, where the last dimension corresponds to the ``channels'' of the convolution.
Intuitively, the convolutional \glspl{NN} compute for each subcarrier and for each \gls{SG} a vector of features.

Next, the sequences of \glspl{SG} corresponding to the hopping patterns are gathered, leading to $K$ tensors denoted by $\Zm^{(1)}_k \in \RR^{S \times 128}$ and $\Zm^{(2)}_k \in \RR^{S \times 128}$, $0 \leq k \leq K-1$, for the first and second \gls{NN}, respectively.
This operation is similar to the one performed for normalizing the received \gls{NN} at preprocessing.
The first \gls{NN} computes for every user $k$ a probability that the user is requesting channel access, denoted by $\widehat{\Pr} \LB A_k \lvert \bar{\Ym} \RB$, by processing $\Zm^{(1)}_k$ with a \gls{MLP}.
Similarly, the second \gls{NN} computes for every user $k$ estimates of the \gls{ToA} and \gls{CFO}, denoted by $\widehat{D}_k$ and $\hat{f}_{\text{off},k}$, from $\Zm^{(2)}_k \in \RR^{S \times 128}$ using two separate \glspl{MLP}.
For both \glspl{NN}, weights sharing is performed across the $K$ hopping patterns for the \glspl{MLP}.

The use of \glspl{MLP} is made possible as short preambles are assumed, leading to small values of $S$.
For example, one preamble repetition corresponds to $S = 4$.
Long preambles could prohibit the use of \glspl{MLP} due to scalability and require a different architecture.
However, as we will see in the next section, the proposed algorithm enables significant gains that could remove the need for longer preambles in many environments.

\subsection{Loss Function}

Training of the \gls{NN} that detects the transmitting users is done on the binary cross-entropy
\begin{equation}
	\label{eq:l1}
	\Lc_1 \coloneqq -\sum_{k=0}^{K-1} \EE \LSB \ln{\widehat{\Pr}\LB A_k \lvert \bar{\Ym} \RB}  \RSB
\end{equation}
which is estimated through Monte Carlo sampling by
\begin{multline}
	\Lc_1 \approx -\frac{1}{B} \sum_{b=1}^{B-1} \sum_{k=0}^{K-1} \Bigg[ A_k^{[b]} \ln{\widehat{\Pr}\LB A_k^{[b]} \lvert \bar{\Ym}^{[b]} \RB}\\
		+ \LB 1 - A_k^{[b]} \RB \ln{1-\widehat{\Pr}\LB A_k^{[b]} \lvert \bar{\Ym}^{[b]} \RB} \Bigg]
\end{multline}
where $B$ is the batch size and the superscript $[b]$ refers to the $b^{\text{th}}$ batch example.

Training of the \gls{NN} that estimates \gls{ToA} and \gls{CFO} is done on the weighted mean squared error
\begin{multline}
	\label{eq:l2}
	\Lc_2 \coloneqq \sum_{k=0}^{K-1} \EE \LSB A_k \text{SNR}_k \LB D_k - \widehat{D}_k \RB^2 \RSB\\
	+ \sum_{k=0}^{K-1} \EE \LSB A_k \text{SNR}_k \LB f_{\text{off},k} - \hat{f}_{\text{off},k} \RB^2 \RSB
\end{multline}
where
\begin{equation}
	\label{eq:snr}
	\text{SNR}_k \coloneqq \frac{\beta_k}{\sigma^2} \frac{1}{S} \sum_{m=0}^{S-1}\abs{H_k\LB \frac{\phi_k[m]}{N} \RB}^2
\end{equation}
is the average \gls{SNR} of user $k$.
Weighting by $A_k$ ensures that only the active users are considered.
Weighting by the \gls{SNR} is motivated by the observation that the errors measured for high \glspl{SNR} are negligible compared to the ones measured for low \glspl{SNR}.
Training on the unweighted MSE therefore results in poor accuracy for high \gls{SNR}.
The loss $\Lc_2$ is estimated by
\begin{multline}
	\Lc_2 \approx \frac{1}{B} \sum_{k=0}^{K-1} \sum_{b=0}^{B-1} A_k^{[b]} \text{SNR}_k^{[b]} \LB D_k^{[b]} - \widehat{D}_k^{[b]} \RB^2\\
	+ \frac{1}{B}\sum_{k=0}^{K-1} \sum_{b=0}^{B-1} A_k^{[b]} \text{SNR}_k^{[b]} \LB f_{\text{off},k}^{[b]} - \hat{f}_{\text{off},k}^{[b]} \RB^2.
\end{multline}
Training of the two \glspl{NN} is jointly performed on the total loss
\begin{equation}
	\Lc \coloneqq \Lc_1 + \Lc_2
\end{equation}
where no weighting is required as $\Lc_1$ and $\Lc_2$ act on different \glspl{NN}.

\begin{figure*}
    \centering

    \begin{tabular}{ccc}
         \begin{subfigure}[t]{0.3\textwidth}
             \centering
             \includegraphics[scale=0.25]{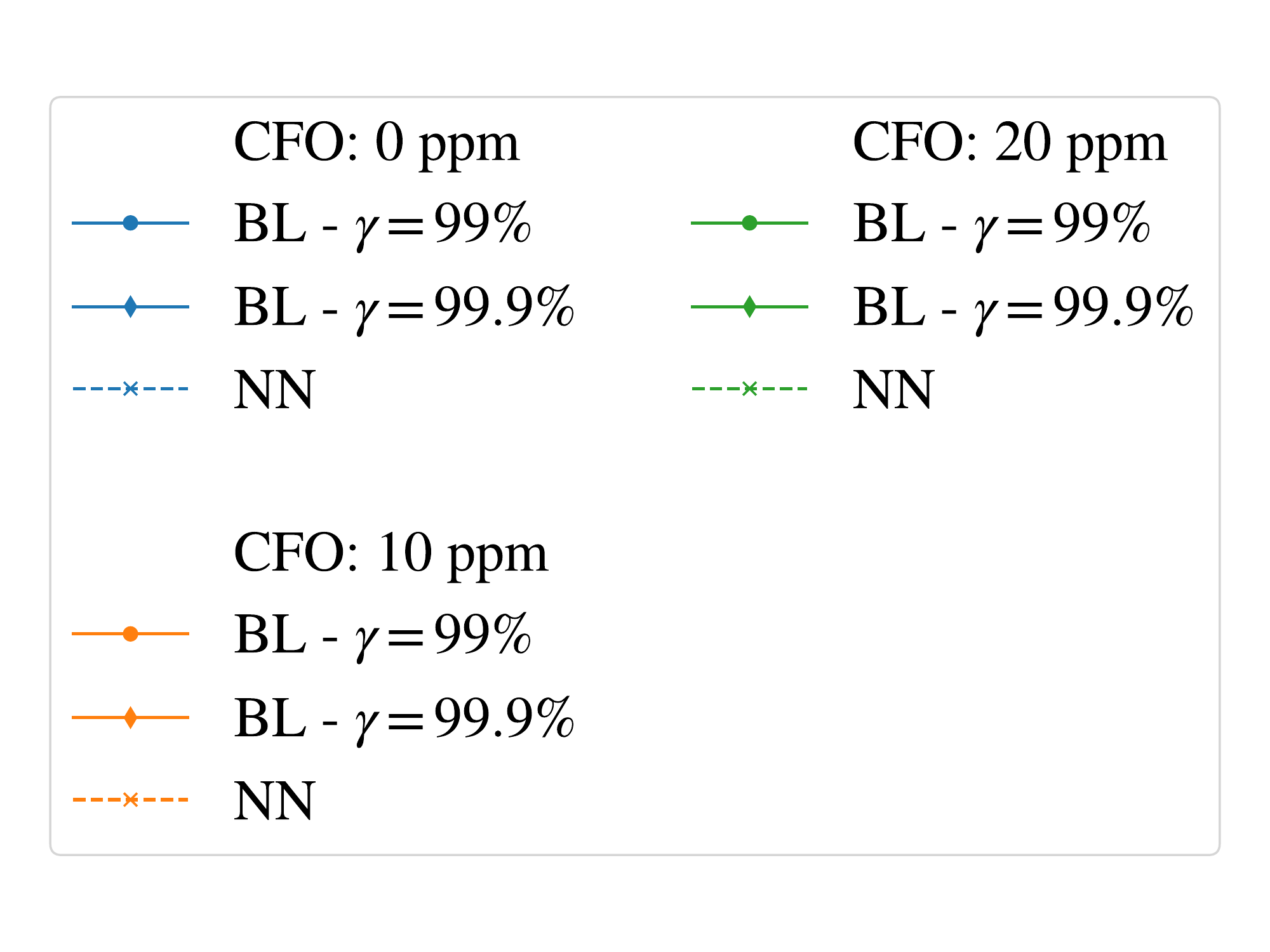}
         \end{subfigure}& 
        \begin{subfigure}[t]{0.3\textwidth}
             \centering
             \includegraphics[scale=0.25]{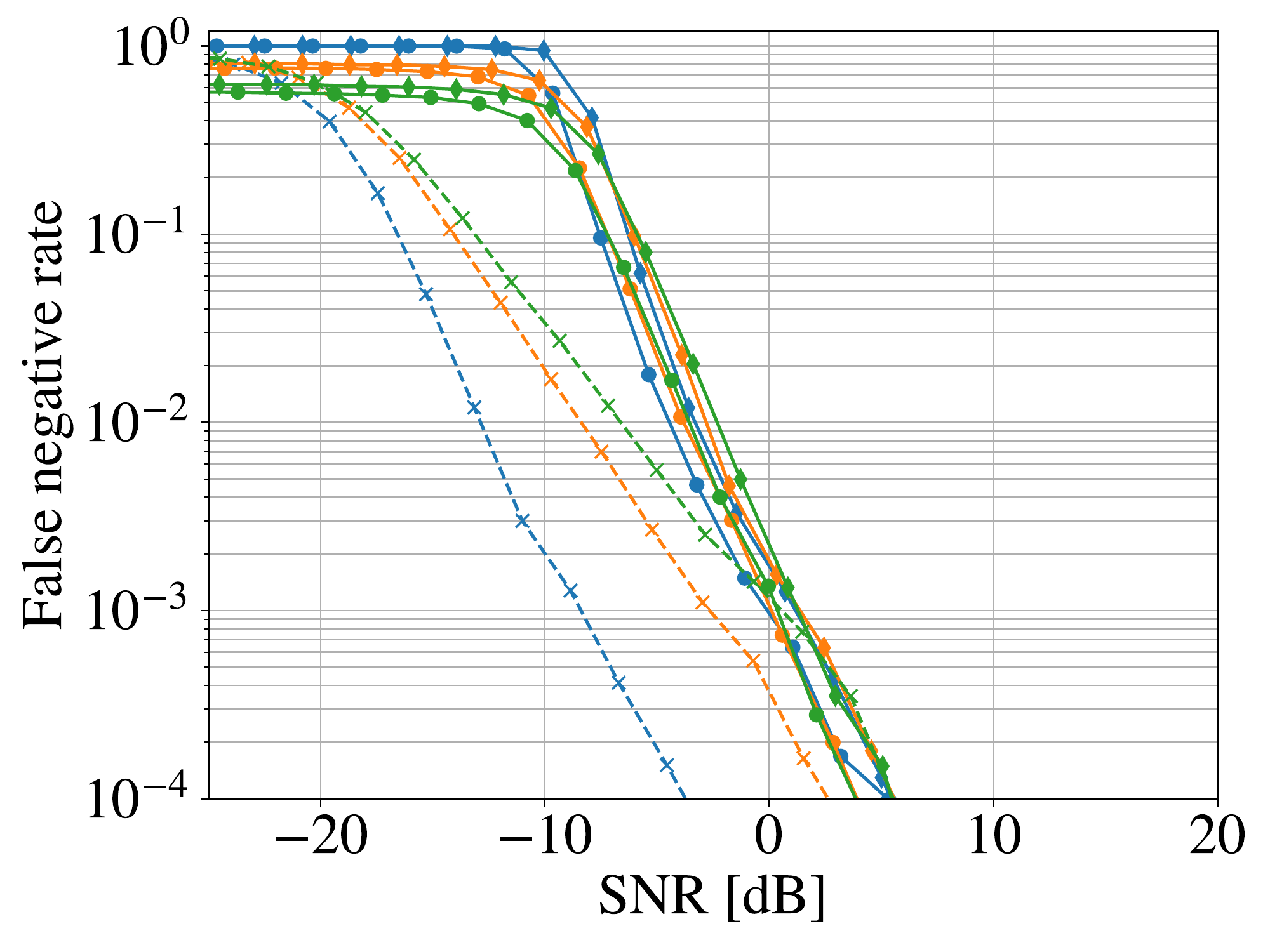}
             \caption{FNR against SNR.\label{fig:fnr_snr}}
         \end{subfigure} &
         \begin{subfigure}[t]{0.3\textwidth}
             \centering
             \includegraphics[scale=0.25]{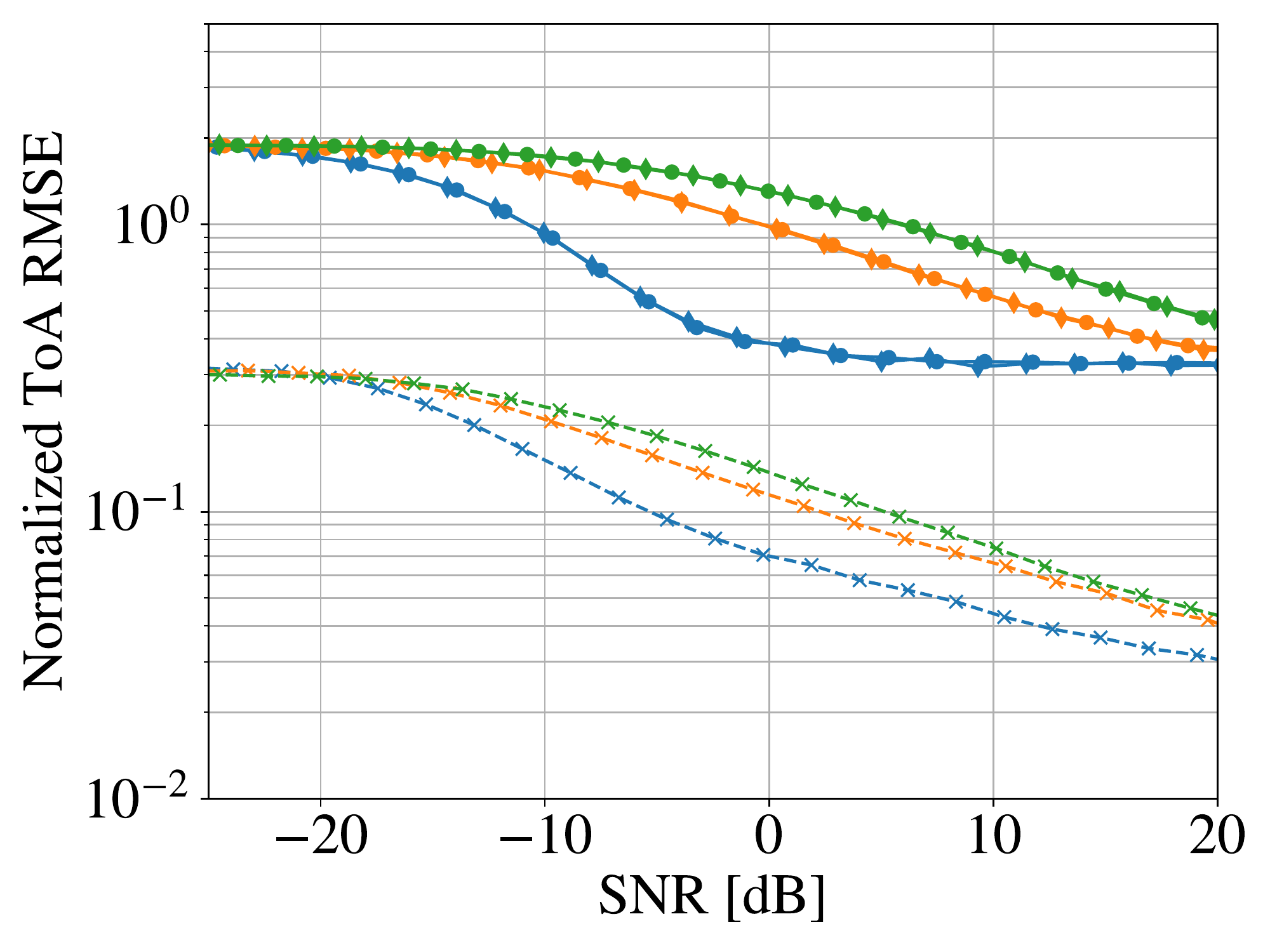}
             \caption{ToA RMSE against SNR.\label{fig:toa_snr}}
         \end{subfigure}\\

         \begin{subfigure}[t]{0.3\textwidth}
             \centering
             \includegraphics[scale=0.25]{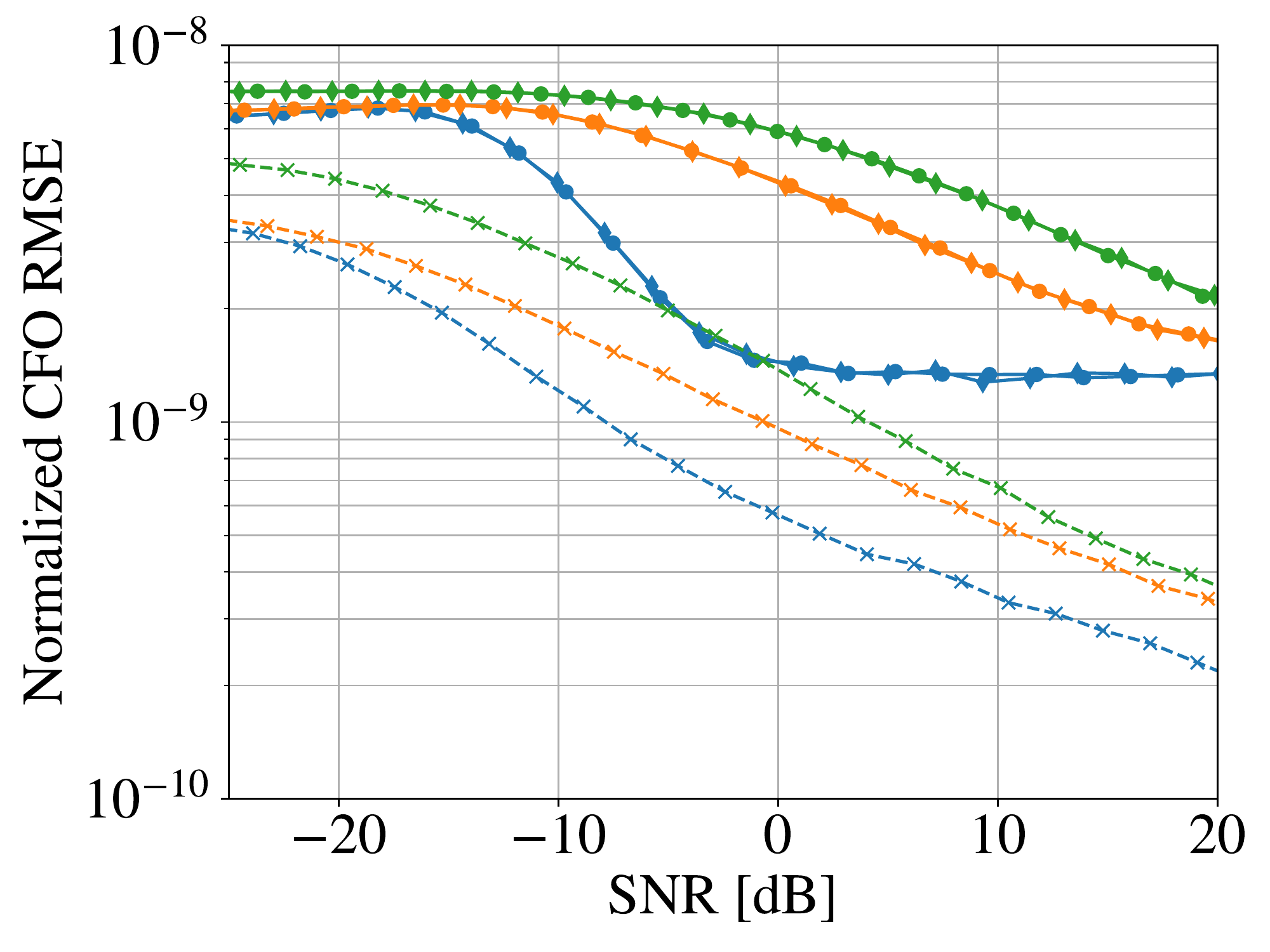}
             \caption{CFO RMSE against SNR.\label{fig:cfo_snr}}
         \end{subfigure} &
         \begin{subfigure}[t]{0.3\textwidth}
            \centering
            \includegraphics[scale=0.25]{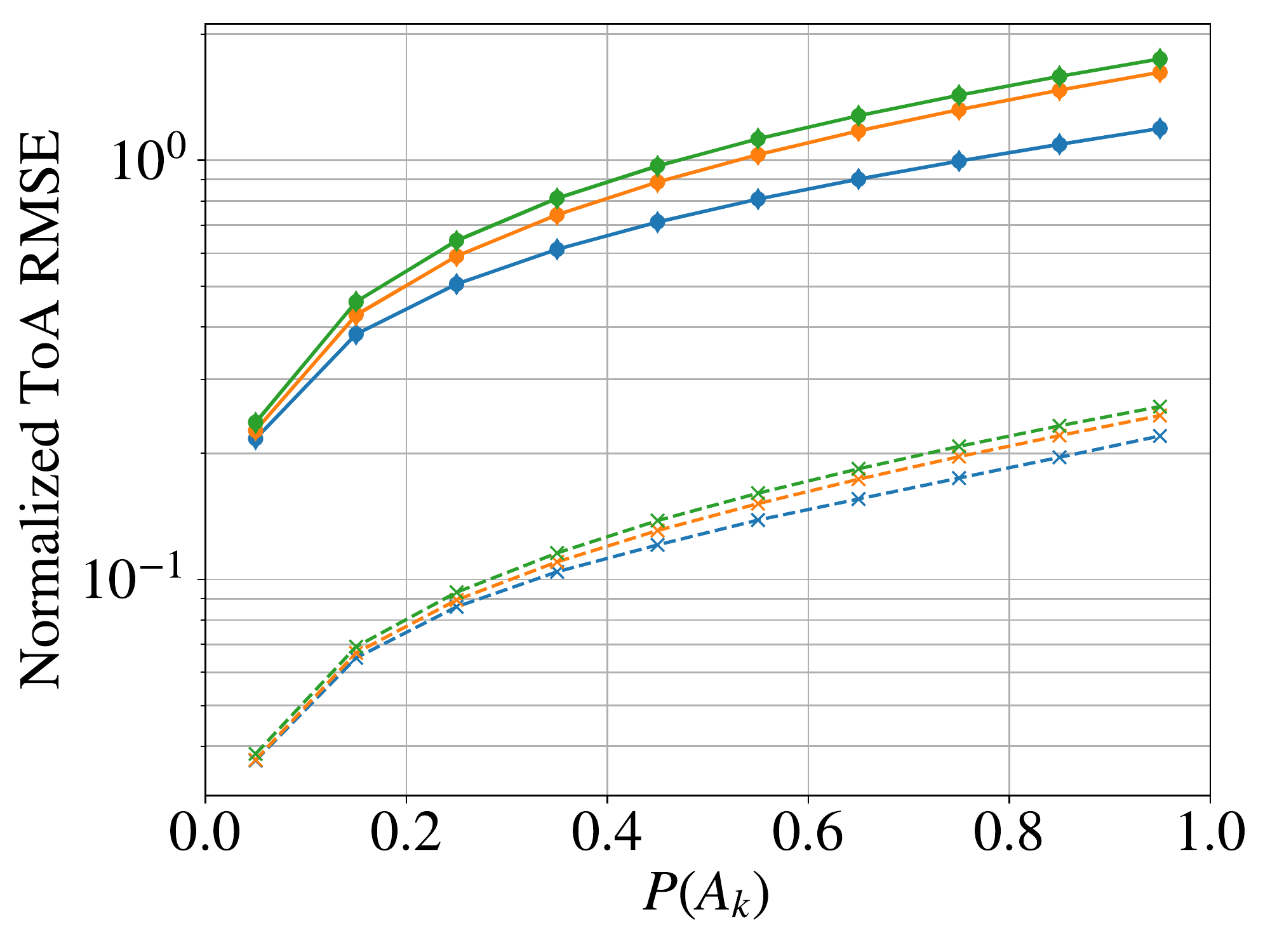}
            \caption{ToA RMSE against $P(A_k)$.\label{fig:toa_ptx}}
         \end{subfigure}&
         \begin{subfigure}[t]{0.3\textwidth}
            \centering
            \includegraphics[scale=0.25]{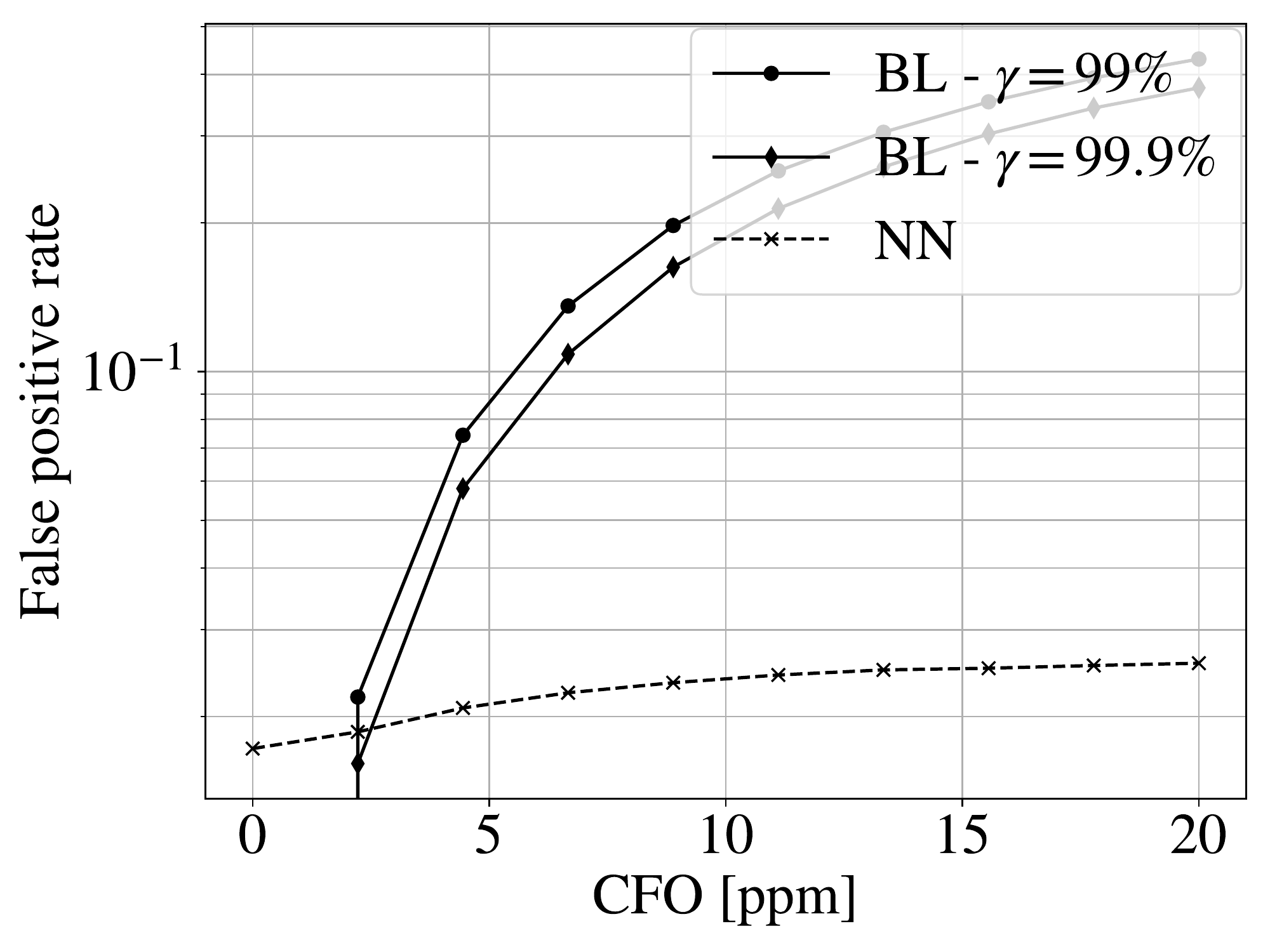}
            \caption{FPR against CFO.\label{fig:fpr_ppm}}
        \end{subfigure}
    \end{tabular}

     \caption{Simulation results. $\gamma$ is the baseline detection threshold. $P(A_k)$ is the probability for a user to transmit.\label{fig:vs_snr}}
\end{figure*}

\section{Simulations Results}

We have benchmarked the previously described algorithm against a state-of-the-art baseline using the Sionna link-level simulator~\cite{hoydis2022sionna}.
The \gls{3GPP} \gls{UMi} channel model~\cite{ts38901} was used, with the carrier frequency set to \SI{3.4}{\giga\Hz} and the sampling frequency set to $\SI{50}{\mega\Hz}$.
The \gls{NPRACH} preamble format 0 was implemented, and the number of repetitions was set to 1, leading to $S = 4$ \glspl{SG}.
Following the \gls{NBIoT} specifications~\cite{ts36211}, the subcarrier spacing $\Delta_f$ was set to \SI{3.75}{\kilo\Hz} and the number of subcarriers allocated to the \gls{NPRACH} to 48.
The maximum number of users that can simultaneously request channel access was set to $K = 48$ with collisionless access.
Note that the proposed method could be trained to detect and resolve collisions.
However, as there is no procedure for handling collisions, such a solution could not be exploited.

Training of the \glspl{NN} was done using the Adam optimizer~\cite{kingma2014adam}, with the batch size set to 64 and the learning rate set to $10^{-3}$.
At training, the probability for a user to request access $P(A_k)$  was independently and uniformly sampled from the range $(0,1)$ for each batch example.
The \gls{CFO}, in \gls{ppm}, of each user and for each batch example was independently and uniformly sampled from the range $(-25,25)$.
Similarly, the \gls{ToA} of each user and for each batch example was independently and uniformly sampled from the range $(0,\SI{66.7}{\mu\s})$, where \SI{66.7}{\mu\s} corresponds to the \gls{CP} length~\cite{ts36211}.
The \glspl{ToA} were added to the path delays generated by the \gls{3GPP} \gls{UMi} channel model.
At both evaluation and training, each batch example consisted of a random drop of $K$ users with randomly chosen large scale parameters to avoid over-fitting to specific channel conditions.
Note that a \emph{single} \gls{NN} was trained over a wide range of transmission probabilities, \glspl{CFO}, and channel conditions, and then evaluated under specific conditions.

To benchmark the proposed method, the synchronization algorithm from~\cite{9263250} was implemented, which builds on previous work~\cite{7569029,8922625}.
This algorithm relies on the configuration of a detection power threshold denoted by $\gamma$, that controls a trade-off between the \gls{FNR} and the \gls{FPR}.
As in~\cite{9263250}, the size of the \gls{FFT} performed by the baseline was set to 256.
Moreover, two values for the detection threshold were used, corresponding to false alarm probabilities of \SI{99.9}{\percent} (as in~\cite{9263250}) and \SI{99}{\percent}.

Fig.~\ref{fig:fnr_snr} shows \gls{FNR} versus \gls{SNR}~\eqref{eq:snr}.
In all figures, the baseline and our approach are referred to as ``BL'' and ``NN'', respectively.
The \gls{CFO} was randomly and uniformly sampled for each user, and the maximum \gls{CFO} value in \gls{ppm} is indicated in the legend.
The probability for a user to transmit was set to $0.5$ for Figures~\ref{fig:fnr_snr}~to~\ref{fig:cfo_snr}
One can see that the \gls{NN}-based method enables gains of up to \SI{8}{\decibel} at an \gls{FNR} of $10^{-3}$ under no \gls{CFO}.
The gains decrease as the \gls{CFO} increases, but remain significant even under high \glspl{CFO} of up to $20\:\textrm{ppm}$.
As expected, the \glspl{FNR} achieved by the baseline are slightly better for the low power detection threshold $\gamma$.
However, this is at the cost of higher \glspl{FPR}, as shown in Fig.~\ref{fig:fpr_ppm}.
Note that Fig.~\ref{fig:fpr_ppm} shows results averaged over all channel realizations, and therefore over all \glspl{SNR}.
The \gls{FPR} is significantly higher with the baseline than with the \gls{NN} approach, except for the lowest \glspl{CFO}.
The steep \gls{FPR} increase observed for the baseline can be explained by the \gls{ICI} caused by the \gls{CFO}~\eqref{eq:rx_freq}, as the energy leaked in adjacent subcarriers erroneously triggers the detection threshold.
The more advanced processing performed by the \gls{NN}-based partially prevents from false detection.

In addition to better detection performance, the \gls{NN}-based algorithm also enables more accurate \gls{ToA} and \gls{CFO} estimation, as shown in Fig.~\ref{fig:toa_snr} and Fig.~\ref{fig:cfo_snr}.
For the baseline, the detection threshold $\gamma$ does not impact the \gls{ToA} and \gls{CFO} estimation accuracy.
Both the baseline and the \gls{NN}-based approach are negatively impacted by increasing \glspl{CFO}, however, the \gls{NN}-based algorithm outperforms the baseline for all \gls{CFO} values.
Moreover, as shown in Fig~\ref{fig:toa_ptx}, the \gls{NN}-based algorithm outperforms the baseline in \gls{ToA} estimation for all transmission probabilities.
Similar results were obtained for \gls{CFO} estimation, but are not shown due to space limitation.

\section{Conclusion}

We have developed a \gls{NN} based solution for \gls{NPRACH} synchronization in \gls{NBIoT}.
We have shown that it enables significant gains in \gls{FNR}, \gls{FPR}, as well as \gls{ToA} and \gls{CFO} estimation accuracy.
These gains are observed for a wide range of \glspl{CFO} and transmission probabilities, showing the robustness of the such an approach.
Our solution is standard-compliant and incurs no additional complexity at the user devices.
It could hence increase battery lifetime by enabling shorter preambles or lower transmit power, which is critical for \gls{NBIoT} applications.

\bibliographystyle{IEEEtran}
\bibliography{IEEEabrv,bibliography}

\end{document}